\documentclass[%
 amsmath,amssymb,
 aps,
prf,
]{revtex4-2}

\usepackage{graphicx}
\usepackage{dcolumn}
\usepackage{bm}
\usepackage{color,graphicx}
\usepackage[dvipsnames]{xcolor}
\usepackage{ulem}
\usepackage{epstopdf, epsfig}
\usepackage{subcaption}
\usepackage{dashrule}
\usepackage{multirow}
\usepackage{natbib,url,hyperref}
\usepackage{amsmath}
\usepackage{lipsum}
\usepackage{tabularx}
\usepackage{rotating}
\usepackage{hyperref}
\usepackage{color}
\usepackage{xcolor}

\definecolor{grey}{rgb}{0.7,0.7,0.7}

\definecolor{cNeutralGray}{RGB}{99,101,105}
\definecolor{cGreenCurrent}{rgb}{0.4660,0.6740,0.1880}
\definecolor{cEverGreen}{rgb}{0,0.5,0}
\definecolor{cCyanCurrent}{rgb}{0.3010, 0.7450, 0.9330}
\definecolor{cBlueSecondary}{RGB}{0, 75, 135}
\definecolor{cViolet}{RGB}{204, 0, 204}
\definecolor{cYellow}{RGB}{255, 255, 51}
\definecolor{cOrangeUtility}{RGB}{242, 160, 0}
\definecolor{cRedUtility}{RGB}{183, 49, 44}

\definecolor{cBlueBrand}{RGB}{47, 126, 178}
\definecolor{cVioletCurrent}{rgb}{0.4940, 0.1840, 0.5560}
\definecolor{cBlack}{rgb}{0, 0, 0}
\definecolor{cYellowCurrent}{rgb}{0.9290, 0.6940, 0.1250}

\makeatletter
\newcommand*\bigcdot{\mathpalette\bigcdot@{.5}}
\newcommand*\bigcdot@[2]{\mathbin{\vcenter{\hbox{\scalebox{#2}{$\m@th#1\bullet$}}}}}
\makeatother

\newcommand{\crline}[1]{ \color{#1} $\bigcdot \,\bigcdot \,\bigcdot \,\bigcdot \,\bigcdot$}

\newcommand{\pizline}[1]{\color{#1} $-\,-\,-$} 
\begin{document}


\title {Sensitivity study of resolution and convergence requirements for\\extended overlap region in wall-bounded turbulence}

\author{Sergio Hoyas}%
\email{serhocal@mot.upv.es}
\affiliation{%
 Instituto Universitario de Matem\'atica Pura y Aplicada, Universitat Polit\`ecnica de Val\`encia, Valencia 46022, Spain {\color{white} a}
}%

\author{Ricardo Vinuesa}%
\email{rvinuesa@mech.kth.se}
\affiliation{%
FLOW, Engineering Mechanics, KTH Royal Institute of Technology, SE-100 44, Stockholm, Sweden
}%

\author{Peter Schmid}
\email{peter.schmid@kaust.edu.sa}
\affiliation{%
King Abdullah University of Science and Technology (KAUST), Thuwal, Saudi Arabia
}%

\author{Hassan Nagib}%
\email{nagib@iit.edu}
\affiliation{%
ILLINOIS TECH (IIT), Chicago, IL 60616, USA
}%

\date{\today}

\begin{abstract}
Direct Numerical Simulations (DNSs) are one of the most powerful tools for studying turbulent flows. Even if achievable Reynolds numbers are lower than those obtained with experimental means, there is a clear advantage since the entire velocity field is known, and any desired quantity can be evaluated. This also includes the computation of derivatives of all relevant terms. One such derivative provides the indicator function, which is the product of wall distance by wall-normal derivative of the mean streamwise velocity.  This derivative may depend on mesh spacing and distribution. However, it is extremely affected by the convergence of the simulation. The indicator function is a cornerstone to understanding inner and outer interactions in wall-bounded flows and describing the overlap region between them. We find a clear dependence of this indicator function on mesh distributions we examined, raising questions about classical mesh and convergence requirements for DNS and achievable accuracy. Within the framework of logarithmic plus linear overlap region, coupled with a parametric study of channel flows, and some pipe flows, sensitivities of extracted overlap parameters are examined and a path is revealed to establishing their high-$Re_\tau$ or near-asymptotic values at modest Reynolds numbers accessible by high quality DNS with reasonable ``cost''.

\end{abstract}


%
\maketitle


Over the last half-century, direct numerical simulation (DNS) has become a widely used approach to study fundamental aspects of wall-bounded turbulence~\citep{hoy06,sim09,lee15,hoy22}. One topic within wall-bounded turbulence that has received extensive attention from the research community due to its significant implications is the overlap region of the mean velocity profile. In this overlap region, the large structures that populate the outer layer co-exist with the smaller ones, which are strongly influenced by the wall and viscous effects. According to the classical literature, the mean velocity profile in this overlap region follows the well-known logarithmic law, with its K\'arm\'an constant, $\kappa$ ~\citep{millikan}:

\begin{equation}\label{eq:001}
    \overline{U_x}^+_{\rm in}(y^+\gg 1) = \frac{1}{\kappa} \ln y^+ + B_0.
\end{equation}

In this equation, $\overline{U_x}$ is the streamwise mean velocity, i.e., we decompose the instantaneous velocity as $U_x=\overline{U_x}+u$. The ``${\rm in}$" subindex (from inner) indicates that we are using the wall characterization of this velocity. The independent variable $y^+$ is the inner-scaled wall-normal coordinate, $y^+ = y u_\tau /\nu$, where $u_\tau$ is the friction velocity and $\nu$ is the fluid kinematic viscosity. Over the years, this has been a topic of active study, and the universality of the log law and the von K\'arm\'an coefficient have been occasionally challenged or reaffirmed; e.g., see references ~\citep{vinuesa_exp,variations,obe22,alc24}. Recently,  Monkewitz and Nagib~\cite{mon23} (MN from now on) shed additional light on this topic, challenging the accumulated knowledge during the last century. MN's main point is to consider in the inner asymptotic expansion a term proportional to the wall-normal coordinate, $\mathcal{O}(Re_{\tau}^{-1})$. Here, $Re_{\tau} = u_\tau \delta / \nu$ is the friction Reynolds number with $\delta$ a characteristic outer length. This $\delta$ can be thought of as the radius $R$ in pipes, the semi-height in channels, $h$, or a length scale related to $\delta_{99}$. Following MN, the pure log law is not observed in channels and pipes (as well as other flows with streamwise pressure gradients) if $Re_{\tau}$ is not above $\mathcal{O}(10^4)$. MN's extended matched asymptotics method reveals an additional term in the expansion of the velocity for this layer, leading to an extra contribution in the overlap layer in the form $S_0 y^+ Re_{\tau}^{-1}$, where $S_0$ is a coefficient. 

Interestingly, the overlap region represented by a combined logarithmic and linear terms exhibits values of the von K\'arm\'an coefficient $\kappa$ consistent with those obtained from skin-friction relations, a fact that suggests that this approach can reveal high-$Re$ trends even at moderate Reynolds numbers. The main point of this matched asymptotic theory is to use a new expression in the overlap region for $\overline{U}_x$. MN's formula reads: 

\begin{equation}\label{eq:010}
\overline{U_x}^+_{\rm in}(y^+\gg 1) \sim \kappa^{-1}\ln y^+ + B_0 + B_1/Re_\tau +S_0  y^+ /Re_\tau.
\end{equation}

\begin{table}
\centering

\begin{tabular}{lcccccccccccccc}
\hline\hline
Case & $ Re_{\tau } $ & $L_{x}/\delta$ & $L_{z}/\delta$ & $\Delta x^{+}$ & $%
\Delta (R\theta ^{+})$ & $\max (\Delta y^{+}$) & $\min (\Delta y^{+})$ & $%
N_{x}$ & $N_{z}$ & $N_{y}$ & ETT & $\varepsilon \times 10^{-4}$ & $\kappa$ & $S_0$ \\ \hline\hline
{\color{blue} PNRF} & 549 & $10\pi$ & $2\pi$ & 5.62 & 3 & 3.2 & 0.018 & 3072 & 1152 & 256 & 
93 & $1.2$ & 0.444 & 2.69  \\ 
{\color{red}  PHRF} & 549 & $10\pi$ & $2\pi$ & 5.62 & 3 & 1.6 & 0.005 & 3072 & 1152 & 512 & 
87 & $2.6$ & 0.449 & 2.88\\ 
{\color{black} PNRC}& 549 & $10\pi$ & $2\pi$ & 11.2 & 4.5 & 3.2 & 0.018 & 1536 & 768 & 256 & 
184 & $0.8$ & 0.451 & 2.77\\ 
{\color{green} PHRC}& 549 & $10\pi$ & $2\pi$ & 11.2 & 4.5 & 1.6 & 0.005 & 1536 & 768 & 512 & 
56 & $3.8$ & 0.429 & 2.53\\ 
{\color{blue} CNRC} & 546  & $8\pi$ & $3\pi$ & 9 & 4.5 & 5.85 & 0.8 & 1536 & 1152 & 251 & 150 & $0.7$ & 0.434 & 1.61 \\ 
{\color{red} CHRC} & 546 & $8\pi$ & $3\pi$ & 9 & 4.5 & 1.68  & 0.05 & 1536 & 1152 & 901 & 150 & $0.3$ & 0.438 & 1.68\\ \hline\hline
\end{tabular}
\caption{DNS case labels and nomenclature of pipe `P' \& channel `C' flows, with `NR' \& `HR' representing classical and high resolution in $y$, and `C' \& ´F' denoting coarse and fine in $x$ and $z$. $L_x$ and $L_z$ are periodic streamwise and spanwise dimensions, and $\delta$ is either channel half height or pipe radius. $\Delta x^{+}$ and $\Delta z^{+}$ are inner-scaled resolutions in terms of dealiased Fourier modes. Wall-normal direction is indicated in both flows by $y$. $N_x, N_z,$ and $N_y$ are numbers of collocation points in three different directions. Time span of simulation is given in terms of eddy turnovers $u_\tau T/\delta$, and $\varepsilon$ is a measure of convergence, defined in \cite{vin16}. Last two columns give value of $\kappa$ and $S_0$ as defined in Equation~(\ref{eq:030}).  Colors in first column are used in Figures~\ref{fig:fig1} (left), \ref{fig:fig3}~\&~\ref{fig:fig4}, and dashed lines for channel.}
    \label{tab:literature}
\end{table}
To obtain accurate values of these coefficients, it is necessary to address two fundamental issues: To find $\kappa$, the challenge is to identify the location and extent of the overlap region, which is weakly dependent on the Reynolds number \cite{mon23}; and To obtain the correct value of $S_0$, it is necessary to obtain with high accuracy the values of ${{\rm d} \overline{U_x}^+}/{\rm d}y^+ $.  It is possible to obtain an equation for the value of $\kappa$ and $S_0$ through the indicator function: 
\begin{equation}\label{eq:020}
 \Xi(y^+) = y^+\frac{{\rm d}\overline{U_x}^+}{{\rm d}y^+}.   
\end{equation}
From equations (\ref{eq:010}) and (\ref{eq:020}), one can obtain the equation for $\kappa$ and $S_0$:

\begin{equation}\label{eq:030}
\Xi(y^+) = \kappa^{-1} +S_0  y^+ /Re_\tau = \kappa^{-1} +S_0 (y/\delta) .
\end{equation}

Thus, to accurately determine $S_0$, it is necessary to compute the indicator function with high accuracy, an aspect which has not been sufficiently investigated in the literature. While a convergence criterion exists for fully developed flows \citep{vin16}, and the domain size of the problem has been examined in depth~\citep{llu18}, the necessary grid spacing has received much less attention. Examples can be found in some very large simulations \citep{hoy22,yao23}, where the authors only report their mesh size and compare their results with those in previous references, which basically do the same. 

\begin{figure}
    \centering
    \begin{subfigure}{0.49\textwidth} 
    \includegraphics[width=0.99\textwidth]{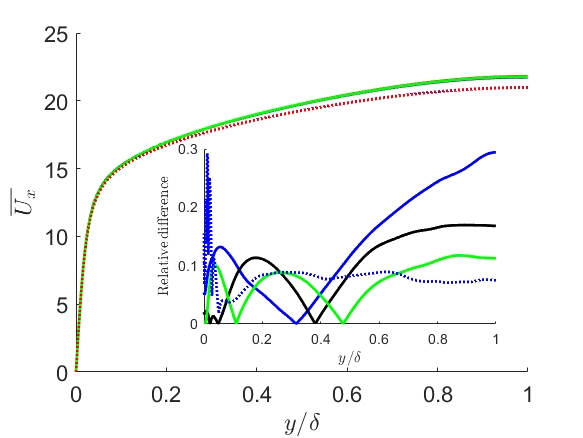}
    \end{subfigure}
    \begin{subfigure}{0.49\textwidth} 
    \includegraphics[width=0.99\textwidth]{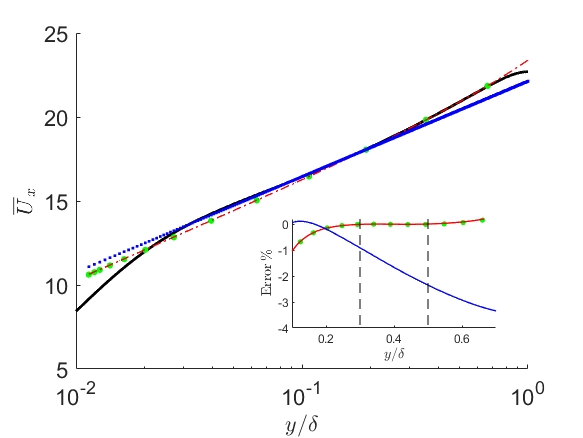}
    \end{subfigure}
        \caption{(Left)  Profile of $\overline{U_x}$ for pipe and channel cases of Table I, with inset showing difference between finest resolution case and others.  (Right).  $\overline{U_x}$ (black line) for case CL83NxNz of Table II, with overlap relations given by Equation~(\ref{eq:001}) (blue dots) and Equation~(\ref{eq:010}) (red dash-dots) \& (green dots), and inset depicting \% error for each.  }
    \label{fig:fig1}
\end{figure}

\begin{table}[]
\begin{tabular}{lcccccccccccccccc}
\hline\hline
Case  & $Re_\tau$ & Line   & $L_x/\delta$ & $L_z/\delta$ & $\Delta_x^+$ & $\Delta_z^+$ & $\max(\Delta y^+)$ & $\min(\Delta y^+)$ & $N_x$ & $N_y$ & $N_z$ & ETT & $\varepsilon \times 10^{-4}$ & $\kappa$ & $S_0$ & ``Cost'' \\ \hline\hline

C21NxNz  & 1000 & \crline{red} & $2\pi$       & $\pi$        & 8.2          & 4.1          & 7.3                & 0.4                                                               & 768   & 383   & 768   & 160 & 2.4                          & 0.386    & 1.27  & 0.08 \\
C42NxNz  & 1000 &   \pizline{red} & $4\pi$     & $2\pi$       & 8.2          & 4.1          & 7.3                & 0.4                                                               & 1536  & 383   & 1536  & 50  & 3.1                          & 0.415    & 1.77  & 0.33 \\
C83NxNz  & 1002 & { \color{black}\hdashrule[0.55ex]{0.8cm}{2pt}{}}	& $8\pi$       & $3\pi$       & 8.2          & 4.1          & 7.3                & 0.4                                                               & 3072  & 383   & 2304  & 47  & 1.5                          & 0.399    & 1.36  & 1    \\
C163NxNz & 1001 &   \pizline{black} & $16\pi$      & $3\pi$       & 8.2          & 4.1          & 7.3                & 0.4                                                               & 6044  & 383   & 2304  & 44  & 0.9                          & 0.401    & 1.46  & 2    \\
C83DxDz  & 1000 & { \color{blue}\hdashrule[0.55ex]{0.8cm}{2pt}{}}	&$8\pi$       & $3\pi$       & 4.1          & 2.0          & 7.3                & 0.4                                                               & 6044  & 383   & 4608  & 29  & 2.8                          & 0.400    & 1.52  & 8    \\
C83NxDz & 1000 &   \pizline{blue}  & $8\pi$       & $3\pi$       & 8.2          & 2.0          & 7.3                & 0.4                                                               & 3072  & 383   & 4608  & 36  & 2.0                          & 0.392    & 1.39  & 2    \\
C83DxNz & 1002 & \crline{blue}   & $8\pi$       & $3\pi$       & 4.1          & 4.1          & 7.3                & 0.4                                                               & 6044  & 383   & 2304  & 15  & 3.3                          & 0.396    & 1.27  & 4    \\
C83HxHz & 1017 &   \pizline{green}  & $8\pi$       & $3\pi$       & 16.4         & 8.2          & 7.3                & 0.4                                                               & 1536  & 383   & 1152  & 83  & 0.2                          & 0.415    & 1.62  & 0.12 \\
C83HxNz  & 1001  &{ \color{green}\hdashrule[0.55ex]{0.8cm}{2pt}{}}	& $8\pi$       & $3\pi$       & 16.4         & 4.1          & 7.3                & 0.4                                                               & 1536  & 383   & 2304  & 37  & 0.8                          & 0.409    & 1.57  & 0.25 \\
C83NxHz& 1018  & \crline{green}  & $8\pi$       & $3\pi$       & 8.2          & 8.2          & 7.3                & 0.4                                                               & 3072  & 383   & 1152  & 10  & 10                           & 0.394    & 1.29  & 0.5  \\
CL83NxNz & 1000 & \crline{black}& $8\pi$       & $3\pi$       & 8.2          & 4.1          & 2.5                & 0.08                                                              & 3072  & 1085  & 2304  & 25  & 9.4                          & 0.394    & 1.42  & 3    \\
C86NxNz & 1001 & ${\color{blue} -\bigcdot-\bigcdot }$ & $8\pi$       & $6\pi$       & 8.2          & 4.1          & 7.3                & 0.4                                                               & 3072  & 383   & 2304  & 35  & 3.4                          & 0.401    &     1.44  & 2 \\\hline\hline
     
\end{tabular}
\caption{Channel-flow DNS cases for higher $Re_\tau$ than in Table I, with nomenclature `Hx,' `Nx,' and `Dx' representing half, classical, and double resolution in $x$, and using same convention for $z$.  Other notation as described for Table I. A theoretical ``cost'' of each simulation compared to case with classical resolution, C83NxNz, is estimated in last column for one ETT.
Colors given in third column are used in Figures~\ref{fig:fig1} (right)~\&~\ref{fig:fig5}.}
\end{table}

We have performed two different numerical experiments to examine this issue and its implications. The details can be found in Table I. On the one hand, we have studied a smooth pipe of radius $R$ and length $L_x=10\pi R$ for $Re_\tau=550$. We will use $y$ to indicate the distance to the wall. We have used four different meshes, as seen in Table 1. The base mesh is the case PLRC with the same resolution in $x$ and $R\Theta$ as a channel. On the other hand, the mesh in the radial direction has twice the number of points needed in a channel due to the skewness of the cells near the pipe center. The details of the mesh can be seen in right side of Figure~\ref{fig:fig1}. Note that for the pipe cases, the wall-normal grid spacing is below the Kolmogorov scale throughot the computational domain. Starting with the PLRC, we doubled the cells in the radial direction (PHRC), azimuthal, and streamwise directions (PLRF), and all of them (PHRF). All these simulations ran for at least 55 eddy-turnover times (ETT), defined as ${\rm ETT}=T u_\tau/R $. 

On the other hand, we have run two sets of channel-flow simulations. The first one (CLRC and CHRC) is a simulation with the same mesh in $x$ and $z$ as in Hoyas and Jim\'enez \citep{hoy06} but different for $y$, as the second one uses a mesh considered enough for $Re_\tau=3000$\citep{alc21b}. The second dataset (Table II) covers different simulations for $Re_\tau=1000$.

Two different codes were employed. OpenpipeFlow was used to run the pipe simulations \cite{yao23}. The code LISO was used for running the channel flow \citep{hoy06,llu21c,yu2022three}. Both codes employ Fourier-decomposition techniques in the streamwise and spanwise directions. Openpipeflow uses a seventh-order finite-difference scheme in the wall-normal direction, while LISO uses a tenth-order compact-finite-difference scheme \cite{llu21c}. 

\begin{figure}
    \centering
    \begin{subfigure}{0.49\textwidth} 
    \includegraphics[width=0.99\textwidth]{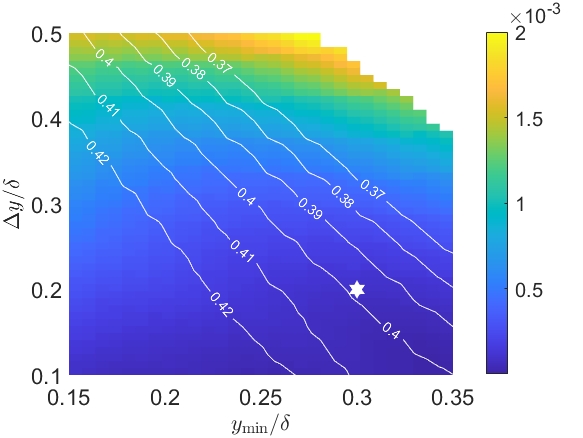}
    \end{subfigure}
    \begin{subfigure}{0.49\textwidth} 
    \includegraphics[width=0.99\textwidth]{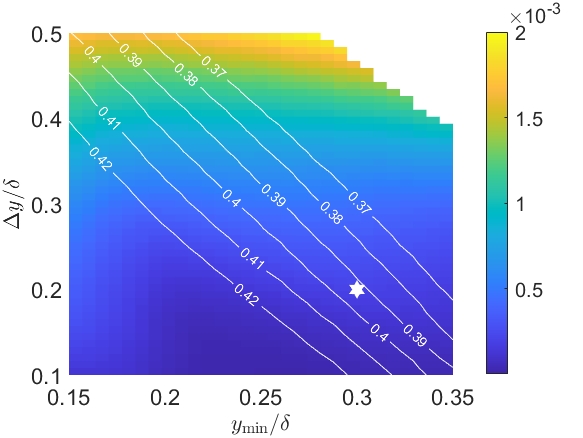}
    \end{subfigure}
    \caption{$L^2$ norm of error between log+lin overlap~(\ref{eq:030}) and DNS data for different intervals. Horizontal axis represents starting point of overlap region, and vertical axis is width of overlap region. White lines are contours of constant $\kappa$. White star indicates selected overlap limits for all other figures. (Left) case C83NxNz. (Right) case CL83NxNz.}
    \label{fig:fig2}
\end{figure}

In the case of the pipe flow, the difference among the four simulations appears small when analyzing the streamwise mean velocity, as displayed in Figure~\ref{fig:fig1}~(left).  Comparable differences are found for channel flows for the mean streamwise velocity as shown in Figure~\ref{fig:fig1}~(right). For a detailed examination of the effects of the DNS resolution in the wall-normal direction and to select the best overlap limits to examine all channel cases, we compare two cases in Figure~\ref{fig:fig2} with one of them using the classical resolutions and the other has only the resolution in wall normal direction doubled. The extracted trends clearly demonstrate the improved trends of results and that the choice of the overlap region $0.3<(y/\delta)<0.5$ is quite appropriate and representative. 

\begin{figure}
    \centering
    \begin{subfigure}{0.49\textwidth} 
    \includegraphics[width=0.99\textwidth]{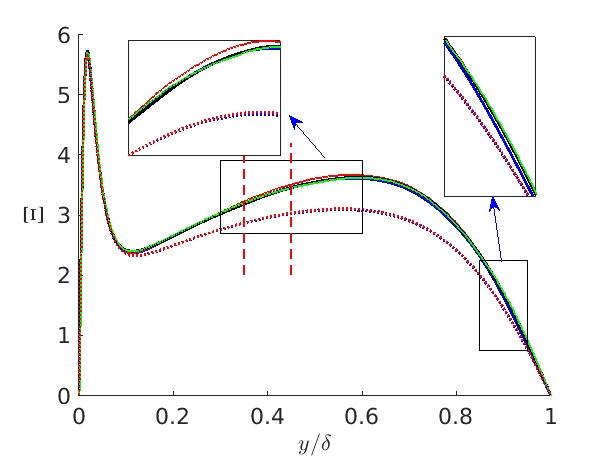}
    \end{subfigure}
    \begin{subfigure}{0.49\textwidth} 
    \includegraphics[width=0.99\textwidth]{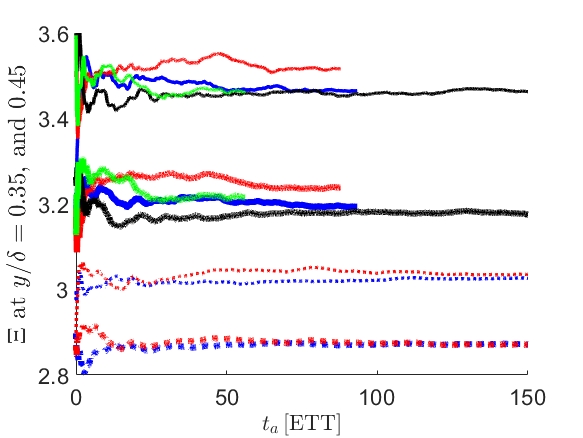}
    \end{subfigure}
        \caption{(Left) Indicator function $\Xi$ for six cases of Table I, depicting differences. (Right) Values of $\Xi$ at $y/\delta=0.35$ and $0.45$ as a function of averaging time $t_a$ expressed in eddy-turnover times (ETT).  Colors as in Table I, with continuous lines for pipes, and dotted lines representing channels. Thick lines, $y/\delta=0.35$. Thin lines, $y/\delta=0.45$.}
     \label{fig:fig3}
     \end{figure}

Part of the motivation to use the indicator function, $\Xi$, is to allow extracting the overlap parameters from experimental data where the accuracy of the measurements is not sufficient, and the density and regular spacing of the data are not adequate for obtaining profile derivatives. For DNS data as demonstrated in the inset of Figure~\ref{fig:fig1} (right), we achieve identical results from the mean velocity profile directly or from its indicator function after differentiation. Even such accuracy is not sufficient for utilizing the indicator function $\Xi$ to extract reliable coefficients of overlap regions. As shown in Figure~\ref{fig:fig3}~(left), the different curves of $\Xi$ do not collapse exactly in the region between $y/\delta=0.3$ and $y/\delta=0.5$, where the slope value of $\Xi$ is critical. This is more clearly appreciated in Figure~\ref{fig:fig3}~(right), where the differences across cases can be more clearly assessed. Notice that the finer mesh, PHRF case, corresponds to the solid red line. The other three cases have not converged to the PHRF's value even after 100 ETT. A better agreement is found in the case of the channels. Here, the convergence at $y/h=0.35$ is obtained after approximately 15 ETT. However, at $y/h=0.45$ the differences are of the order of 2\% after 80 ETT, which is an extremely lengthy computation for high-Reynolds-number simulations. 

\begin{figure}
 \centering
        \begin{subfigure}{0.49\textwidth} 
    \includegraphics[width=0.95\textwidth]{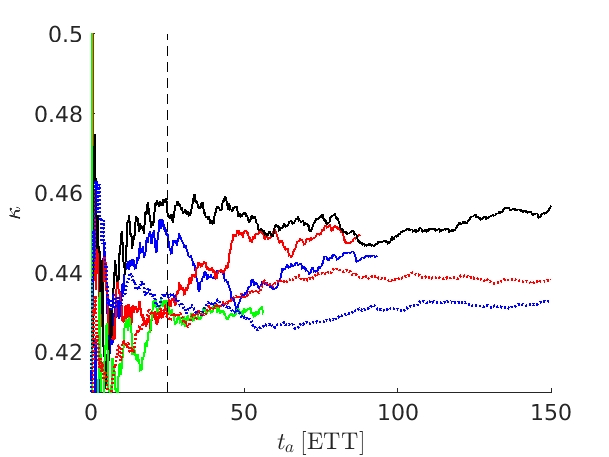}
    \end{subfigure}
    \begin{subfigure}{0.49\textwidth} 
    \includegraphics[width=0.95\textwidth]{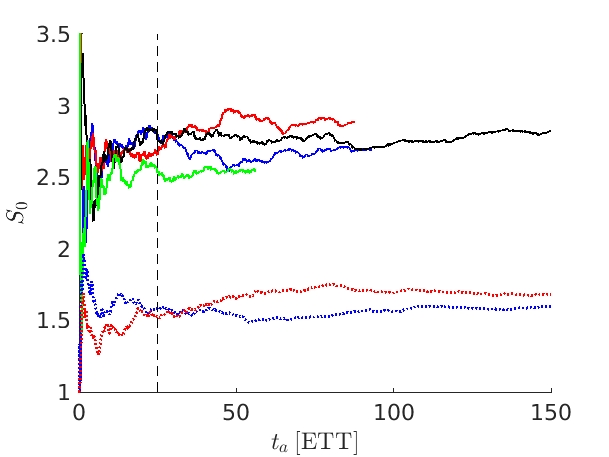}
    \end{subfigure}
    \caption{ Extracted overlap parameters $\kappa$ (Left) and $S_0$ (Right) for six cases of Table I as functions of averaging time $t_a$ expressed in ETT. Colors as in Table I, with continuous lines for pipes, and dotted lines representing channels.}
    \label{fig:fig4}

\end{figure}

The effect of this lack of adequate resolution can be better evaluated by the time-averaged values of the overlap coefficients $\kappa$, $B$, and $S_0$ as functions of ETT. After 50 ETTs, the differences among the cases are still above 3$\%$, as shown in Figure~\ref{fig:fig4}. Again, the convergence for the channel case is better, but there remains a small difference up to ETT of 150, which corresponds to very expensive computations. In any case, at least 20 ETTs are needed to obtain some level of convergence. Such differences in the indicator function $\Xi$ should be smaller for any reliable simulation to predict the values of $\kappa$, $B$, and $S_0$. 

Finally, in the case of pipe flow, the values of these coefficients appear to depend on the azimuthal resolution, raising questions about the mesh aspect ratio of the grids. A careful study of such mesh-variation effects in flows near their singularities, as in case around the axis of pipes, is highly recommended. Such a study should also examine the sensitivity of the indicator function $\Xi$ to the order of the finite difference scheme used in the DNS. In general, the results demonstrate that classical and commonly used mesh sizes in the wall-normal direction for DNSs are not sufficiently fine to extract coefficients of the overlap region in wall-bounded turbulent flows, such as $\kappa$ and $S_0$ with good accuracy.


\begin{figure}
 \centering
        \begin{subfigure}{0.49\textwidth} 
    \includegraphics[width=0.99\textwidth]{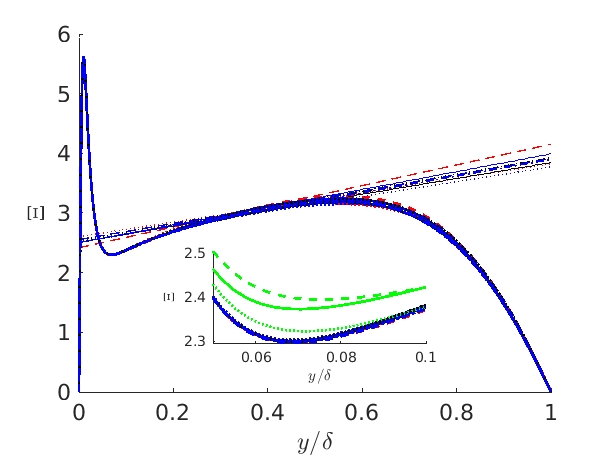}
    \end{subfigure}
    \begin{subfigure}{0.49\textwidth} 
    \includegraphics[width=0.99\textwidth]{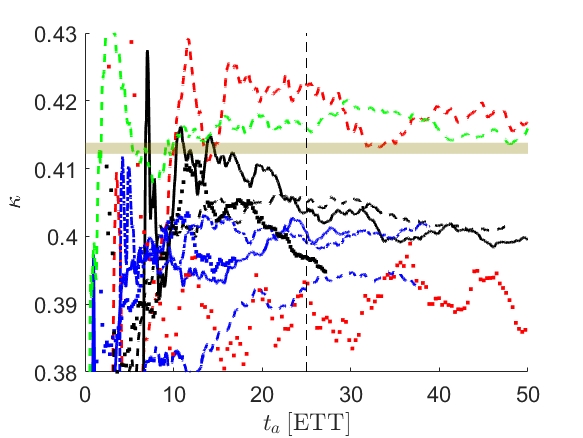}
    \end{subfigure}
    \caption{(Left) Indicator function $\Xi$ for cases of Table II, with straight lines depicting log+lin overlap relation of Equation~(\ref{eq:030}) for each case, and inset depicting the limitations in cases with worst resolution by failing to capture near-wall minimum value. (Right) Range of extracted $\kappa$'s as function of ETT for cases displayed with line types \& colors shown in Table II, with best resolved cases for $Re_\tau$~$\approx$~$1000$ yielding $\kappa$~$\approx$~$0.4$ \& $S_0$~$\approx$~$1.5$. Extracted parameters of Equation~(\ref{eq:030}) for ``converged'' higher $Re_\tau$~$\approx$~$5200$ is shown by wide yellow line: $\kappa$~$\approx$~$0.413$ \& $S_0$~$\approx$~$1.15$~\cite{nag24,lee15}.}
    \label{fig:fig5}
\end{figure}

In conclusion, DNSs still stand as a powerful tool for exploring turbulent flows. However, the accuracy of DNS to compute certain important derivatives, such as those required for the indicator function $\Xi$, is challenged. To examine the sensitivities of DNS to various simulation parameters by accessible values of $Re_\tau$'s while allowing a multitude of cases in a parametric study, we extended the cases of Table I to the $Re_\tau~\approx~1000$ cases of Table II. In every case, the results of the turbulence intensities are below a difference of 1\% for either normal or double resolution.

However, for calculating $\Xi$, accurate values of the wall-normal derivative of the streamwise velocity must be established. Our investigation reveals that the indicator function exhibits sensitivity around the classic mesh size commonly used in DNSs. In particular, several coefficients of the overlap region required for the Monkewitz and Nagib’s model of wall-bounded flows under streamwise pressure gradients \cite{mon23} cannot be found with less than $3\%$ difference among the best-resolved meshes and nearly-converged cases we tested. Differences of as much as $8\%$ are found between all the cases of Table II, including some with inadequate resolution or convergence or different box sizes. For example, this creates a new uncertainty source in the estimation of the von K\'arm\'an coefficient that is very important for modeling and predicting wall-bounded turbulent flows, as highlighted recently by Monkewitz's careful evaluation \cite{mon24}. 

Such uncertainty extends with different levels to other quantities needing wall-normal derivatives, including the normal stresses, as we have also examined recently. The indicator functions, $\Xi$, displayed in Figure 1 of \cite{nag24} for DNS of channel flows with $550<Re_\tau<16,000$, demonstrate that the trend with Reynolds number is for $\kappa$ to continue to increase beyond the value represented by the wide yellow line in Figure~\ref{fig:fig4}.  Beyond $Re_\tau$ of around $5,000$ the effects on the extracted parameters may not be only that due to Reynolds number. It may also include effects of resolution, convergence, computational box size, and order of the numerical method. From the comparisons of indicator functions of mean velocity profiles and streamwise normal stresses in \cite{nag24}, it appears that the mean flow sensitivity is more than that for the turbulence field. Considering the added limitations and uncertainties with some experimental measurements, far more attention is required to validate models and the extracted parameters.  The log+lin model of MN~\cite{mon23} compared to the pure-log model, has allowed evaluating the sensitivities of their parameters examined here and permitted the extraction of their high-$Re_\tau$ or near-asymptotic values at modest Reynolds numbers accessible by high quality DNS with reasonable ``cost''.\\

\textbf{Data availability:} The data used for this paper can be obtained by contacting S. Hoyas at serhocal@mot.upv.es. 

\begin{acknowledgments}
For computer time, this research used the resources of the Supercomputing Laboratory at King Abdullah University of Science \& Technology (KAUST) in Thuwal, Saudi Arabia, Project K1652. RV acknowledges the financial support from ERC grant no. `2021-CoG-101043998, DEEPCONTROL'. Views and opinions expressed are however those of the author(s) only and do not necessarily reflect those of European Union or European Research Council. Neither the European Union nor granting authority can be held responsible for them.  SH is  funded by project PID2021-128676OB-I00 by Ministerio de Ciencia, innovación y Universidades / FEDER. 
\end{acknowledgments}

\bibliography{turbulence}

\end{document}